\documentclass[12pt]{article}

\usepackage{graphicx}
\begin{document}

\begin{center}
{\bf Magnetic black holes with generalized ModMax model of nonlinear electrodynamics} \\
\vspace{5mm} S. I. Kruglov
\footnote{E-mail: serguei.krouglov@utoronto.ca}
\underline{}
\vspace{3mm}

\textit{Department of Physics, University of Toronto, \\60 St. Georges St.,
Toronto, ON M5S 1A7, Canada\\
Department of Chemical and Physical Sciences, University of Toronto,\\
3359 Mississauga Road North, Mississauga, Ontario L5L 1C6, Canada} \\
\vspace{5mm}
\end{center}
\begin{abstract}
Recently Bandos, Lechner, Sorokin, and Townsend [Phys. Rev. D \textbf{102}, 121703 (2020)] proposed Modified Maxwell (ModMax) model of nonlinear duality-invariant conformal electrodynamics. Here, Generalized ModMax (GenModMax) model of nonlinear electrodynamics coupled to general relativity is studied. The metric and mass functions, and their asymptotic as $r\rightarrow\infty$ and $r\rightarrow 0$ of a magnetic black hole are obtained. Corrections to the Reissner--Nordstr\"{o}m solution are found and we show that for some model parameters the black hole is regular. The Hawking temperature and heat capacity of black holes are calculated and phase transitions are investigated. We demonstrate that black holes are not stable for certain model parameters.
\end{abstract}

\section{Introduction}

The first model of nonlinear electrodynamics (NED) was proposed by Born and Infeld (BI) \cite{Born}. The Maxwell electrodynamics is dual and scale invariant. In BI electrodynamics
the electric field at the center of point-like particles and the self-energy of charges are finite. The BI electrodynamics appears in  string
theory at low energies \cite{Fradkin, Tseytlin}. It was demonstrated that NED models \cite{Shabad, Shabad1, Kr1, Kr2, Kr3} also possess similar attractive features as BI electrodynamics. In quantum electrodynamics due to loop corrections \cite{Heisenberg, Schwinger, Adler} nonlinear terms are generated and the dual and scale symmetries are broken. In ModMax model of nonlinear electrodynamics \cite{Bandos} the duality and conformal symmetries hold. Some applications of ModMax electrodynamics were investigated in \cite{Kosyakov, Habib, Bandos1,Townsend, Flores,Kubiznak}.
Generalized ModMax (GenModMax) electrodynamics with four parameters was proposed in \cite{Kruglov}. In the weak-field limit and small $\gamma$-parameter GenModMax model becomes ModMax electrodynamics. At some parameters GenModMax model is converted in BI-type model \cite{Kruglov0, Kruglov1}, the generalized BI model \cite{Kruglov2}, and  BI model \cite{Born}. Here, we obtain black hole (BH) solutions and investigate BH thermodynamics in the framework of GenModMax model. Black holes were studied in many papers, e.g. \cite{Bardeen, Ayon1, Bronnikov, Breton, Hayward, Breton1, Lemos, Lemos1, Balart, Kruglov3, Kruglov4, Kruglov5, Frolov}.
It is worth mentioning that a general pure electric solution with the general Lagrangian $L({\cal F})$ was presented by Pellicer and
Torrence \cite{Pellicer}. A construction of exact black hole solutions with electric or magnetic charges in General Relativity was considered in \cite{Fan} (see also a comment in \cite{Bronnikov1}), and a pure magnetic solution was given in \cite{Bronnikov}.

The paper is organised as follows. In section 2 the GenModMax model is introduced.  We study
GenModMax electrodynamics coupled to gravity in section 3. The magnetically charged black hole is considered.
The asymptotic of the metric and mass functions as $r\rightarrow\infty$ and $r\rightarrow 0$ are obtained. We find corrections to the Reissner--Nordstr\"{o}m (RN) solution. The Hawking temperature and heat capacity are calculated in section 4. It is demonstrated that BHs are not stable for some parameters and there are phase transitions. Section 5 is a conclusion.

The units with $c=\hbar=1$ are used and the metric signature is $\eta=\mbox{diag}(-1,1,1,1)$.

\section{A model of GenModMax electrodynamics}

GenModMax electrodynamics is described by the Lagrangian \cite{Kruglov}
\begin{equation}
{\cal L}=\frac{1}{\beta}\left(1-\left(1+\frac{\beta [{\cal F}\cosh(\gamma)-\sqrt{{\cal F}^2+{\cal G}^2}\sinh(\gamma)]}{\sigma}-\frac{\beta \lambda{\cal G}^2}{2\sigma}\right)^\sigma\right),
\label{1}
\end{equation}
and the Lagrangian of ModMax electrodynamics is given by
\begin{equation}
L=-{\cal F}\cosh(\gamma)+\sqrt{{\cal F}^2+{\cal G}^2}\sinh(\gamma),
\label{2}
\end{equation}
with
\begin{equation}
{\cal F}=\frac{1}{4}F_{\mu\nu}F^{\mu\nu}=\frac{1}{2}\left(\textbf{B}^2-\textbf{E}^2\right),~~~~{\cal G}=\frac{1}{4}F_{\mu\nu}\tilde{F}^{\mu\nu}=\textbf{B}\cdot \textbf{E}.
\label{3}
\end{equation}
The parameters $\beta$ and $\lambda$ have the dimensions of (length)$^4$, and $\sigma$ and $\gamma$ are the dimensionless parameters,
$F_{\mu\nu}=\partial_\mu A_\nu-\partial_\nu A_\mu$ is the field strength tensor, and $\tilde{F}^{\mu\nu}=(1/2)\epsilon^{\mu\nu\alpha\beta}F_{\alpha\beta}$
is the dual tensor. At $\beta L\ll 1$, $\beta\lambda{\cal G}^2\ll 1$ the Lagrangian (1) gives
\begin{equation}
{\cal L}=L+\frac{\beta(1-\sigma)L^2}{2\sigma} + \frac{\lambda}{2}{\cal G}^2+{\cal O}(LG^2)+
{\cal O}(L^3).
\label{4}
\end{equation}
Thus, in the weak-field limit and small $\gamma$ Lagrangian (1) approaches to the ModMax Lagrangian (2). The Heisenberg--Euler-type electrodynamics \cite{Kruglov6} is reproduced at $\beta L\ll 1$, $\beta\lambda{\cal G}^2\ll 1$ and $\gamma=0$.
The asymptotic of the electric field ($\sigma<1$) as $r\rightarrow 0$ and $r\rightarrow\infty$
were obtained for the flat space-time and spherical symmetry, and are given by \cite{Kruglov}
\[
E=\sqrt{\frac{2\sigma}{\beta}}\exp\left(-\frac{\gamma}{2}\right)-\frac{(2\sigma)^{(2-\sigma)/(2(1-\sigma))}r^{2/(1-\sigma)}}
{2\beta^{(2-\sigma)/(2(1-\sigma))}q^{1/(1-\sigma)}}\exp\left(\frac{\gamma\sigma}{2(1-\sigma)}\right)
\]
\[
+{\cal O}(r^{4/(1-\sigma)})~~~~~~r\rightarrow 0,
\]
\begin{equation}
E=\frac{q}{r^2}\exp(-\gamma)-\frac{(1-\sigma)\beta q^3}{2\sigma r^6}\exp(-2\gamma)+{\cal O}(r^{-8})~~~~~~r\rightarrow\infty.
\label{5}
\end{equation}
Equation (5) shows the correction to Coulomb's law as $r\rightarrow\infty$.
At $\sigma=1$, $\lambda=0$, $\gamma=0$, one has Maxwell’s electrodynamics and we arrive at the Coulomb law $E=q/r^2$ as $r\rightarrow\infty$.
It follows from Eq. (5) that the electric field of the point-like charged particle in the center is finite and has the maximum value \cite{Kruglov}
\begin{equation}
E(0)=\sqrt{\frac{2\sigma}{\beta}}\exp\left(-\frac{\gamma}{2}\right).
\label{6}
\end{equation}
As a result, the electric field is damped due to parameter $\gamma$. The energy-momentum tensor of GenModMax electrodynamics is given by
\begin{equation}
T_{\mu\nu}=F_{\mu\alpha}\left({\cal L}_{\cal F}F_{~\nu}^{\alpha}+{\cal L}_{\cal G}\tilde{F}_{~\nu}^{\alpha}\right)-g_{\mu\nu}{\cal L},
\label{7}
\end{equation}
giving the energy density
\begin{equation}
T_{0}^{~0}=-E^2{\cal L}_{\cal F}-G{\cal L}_{\cal G}-{\cal L},
\label{8}
\end{equation}
where ${\cal L}_{\cal F}=\partial {\cal L}/\partial {\cal F}$, ${\cal L}_{\cal G}=\partial {\cal L}/\partial {\cal G}$. It should be noted that ModMax model and the two-parametric generalized BI model at $\sigma=1/2$, $\lambda=\beta$ are duality invariant \cite{Bandos1, Kruglov}. The conformal invariance ($T_\mu^{~\mu}=0$) takes place only for $\sigma=1$, $\lambda=0$ corresponding to the ModMax model.

\section{Magnetically charged black hole}

The action of GenModMax electrodynamics coupled to general relativity is given by
\begin{equation}
S=\int d^4x\sqrt{-g}\left(\frac{1}{2\kappa^2}R+ {\cal L}\right),
\label{9}
\end{equation}
where $\kappa^2=8\pi G\equiv M_{Pl}^{-2}$, $G$ is Newton's constant, $M_{Pl}$ is the reduced Planck mass, and $R$ is the Ricci scalar.
 The Einstein equation and electromagnetic field equations follow from Eq. (9)
\begin{equation}
R_{\mu\nu}-\frac{1}{2}g_{\mu\nu}R=-\kappa^2T_{\mu\nu},
\label{10}
\end{equation}
\begin{equation}
\partial_\mu\left[\sqrt{-g}{\cal L}_{\cal F}F^{\mu\nu}\right]=0.
\label{11}
\end{equation}
 The line element with the spherical symmetry is
\begin{equation}
ds^2=-f(r)dt^2+\frac{1}{f(r)}dr^2+r^2(d\vartheta^2+\sin^2\vartheta d\phi^2).
\label{12}
\end{equation}
We consider the magnetically charged BH because it can be nonsingular compared to electrically charged BH \cite{Bronnikov}.
For a pure magnetic black holes the invariant ${\cal G}$ vanishes and we use methods obtained for $L({\cal F})$.
The metric function $f(r)$ is defined by the relation \cite{Bronnikov}
\begin{equation}
f(r)=1-\frac{2GM(r)}{r},
\label{13}
\end{equation}
where the mass function is given by
\begin{equation}
M(r)=\int_0^r\rho_M(r)r^2dr=m_M-\int^\infty_r\rho_M(r)r^2dr,
\label{14}
\end{equation}
where the BH magnetic mass is given by $m_M=\int_0^\infty\rho_M(r)r^2dr$ and $\rho_M$ is the magnetic energy density. The BH mass possesses the electromagnetic nature. The magnetic energy density found from Eq. (8) for $E=0$ is
\begin{equation}\label{15}
  \rho_M=T_0^{~0}=\frac{1}{\beta}\left[\left(1+\frac{\beta B^2\exp(-\gamma)}{2\sigma}\right)^\sigma-1\right].
\end{equation}
The magnetic induction field of the magnetic monopole is $B=Q/r^2$ and ${\cal F}=Q^2/(2r^4)$, where $Q$ is a magnetic charge. Thus, we
consider a black hole as a magnetic monopole.
From Eqs. (14) and (15) we find the mass function
\begin{equation}\label{16}
  M(u)=m_M+\frac{Q^{3/2}u^3}{3(2\sigma)^{3/4}\beta^{1/4}\exp(3\gamma/4)}\left[
_2F_1\left(-\frac{3}{4},-\sigma;\frac{1}{4};-\frac{1}{u^4}\right)-1\right],
\end{equation}
where $_2F_1(a,b;c;z)$ is the hypergeometric function and we introduced the dimensionless variable
\begin{equation}\label{17}
u=\frac{r(2\sigma)^{1/4}\exp(\gamma/4)}{\beta^{1/4}\sqrt{Q}}.
\end{equation}
Introducing the dimensionless magnetic mass of the BH
\begin{equation}\label{18}
\bar{m}_M=\beta^{1/4}Q^{-3/2}\exp(3\gamma/4)m_M
\end{equation}
we calculate $\bar{m}_M$ represented in Table 1.
%$\chi\equiv(2\sigma)^{3/4}\beta^{1/4} m_M/q^{3/2}$
\begin{table}[ht]
\caption{The dimensionless magnetic mass of the BH}
\centering
\begin{tabular}{c c c c c c c c c }\\[1ex]
% centered columns
\hline
$\sigma$ & 0.1 & 0.2 & 0.3 & 0.4 & 0.5 & 0.6 & 0.7  \\[0.5ex]
\hline
$\bar{m}_M$ & 0.525 & 0.667 & 0.806 & 0.977 & 1.236 & 1.774 & 4.278  \\[0.5ex]
\hline
\end{tabular}
\end{table}
With the help of Eqs. (13) and (14) we obtain the metric function
\begin{equation}\label{19}
  f(r)=1-\frac{2Gm_M}{r}-\frac{2Gr^2}{3\beta}\left[
_2F_1\left(-\frac{3}{4},-\sigma;\frac{1}{4};-\frac{\beta Q^2\exp(-\gamma)}{2\sigma r^4}\right)-1\right].
\end{equation}
It is convenient to introduce the dimensionless parameter
\begin{equation}\label{20}
  b=\frac{\sqrt{\beta\sigma}\exp(\gamma/2)}{\sqrt{2}GQ}.
\end{equation}
Then the metric function (19) becomes
\begin{equation}\label{21}
  f(u)=1-\frac{(2\sigma)^{3/4}\bar{m}_M}{bu}-\frac{u^2}{3b}\left[
_2F_1\left(-\frac{3}{4},-\sigma;\frac{1}{4};-\frac{1}{u^4}\right)-1\right].
\end{equation}
Now we use the asymptotic of the hypergeometric function which converges at $|z|\leq 1$ \cite{Bateman}
\begin{equation}\label{22}
_2F_1\left(a,b;c;z\right)=\sum_{n=0}^\infty\frac{(a)_n(b)_nz^n}{(c)_nn!},
\end{equation}
where $(q)_0=1$, $(q)_n=q(q+1)...(q+n-1)$, $n=1, 2, 3,...$.
Making use of Eq. (22) we obtain the asymptotic of the hypergeometric function as $r\rightarrow\infty$
\begin{equation}\label{23}
  _2F_1\left(-\frac{3}{4},-\sigma;\frac{1}{4};-\frac{1}{u^4}\right)=1-
\frac{3\sigma}{u^4}+\frac{3\sigma(1-\sigma)}{10u^8}-\frac{\sigma(1-\sigma)(2-\sigma)}{18u^{12}}+{\cal O}(u^{-16}).
\end{equation}
From Eqs. (20), (21) and (23) one finds the asymptotic of the metric function as $u\rightarrow\infty$ ($r\rightarrow\infty$)
\[
  f(r)=1-\frac{2Gm_M}{r}+\frac{GQ^2}{r^2}\exp(-\gamma)-\frac{(1-\sigma)G\beta Q^4}{20\sigma r^6}\exp(-2\gamma)
\]
\begin{equation}\label{24}
+\frac{(1-\sigma)(2-\sigma)G\beta^2 Q^6}{216\sigma^2 r^{10}}\exp(-3\gamma)+{\cal O}(r^{-14}).
\end{equation}
Equation (24) shows corrections to RN solution in the order of ${\cal O}(r^{-6})$.
As $r\rightarrow \infty$ we have $f(\infty)=1$ and the space-time becomes flat. At $\sigma=1$ ($E=0$) one has ModMax electrodynamics and we arrive from Eq. (24) at the RN solution.
The plots of the function $f(x)$ are represented in Figs. 1, 2 and 3.
\begin{figure}[h]
\includegraphics[height=3.0in,width=3.0in]{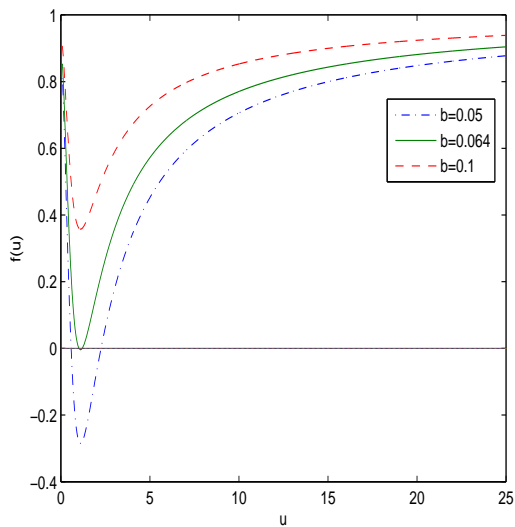}
\caption{\label{fig.1}The plot of the function $f(u)$  for $\sigma=0.1$. Dashed-dotted line corresponds to $b=0.05$, solid line corresponds to $b=0.064$ and dashed line corresponds to $b=0.1$.}
\end{figure}
\begin{figure}[h]
\includegraphics[height=3.0in,width=3.0in]{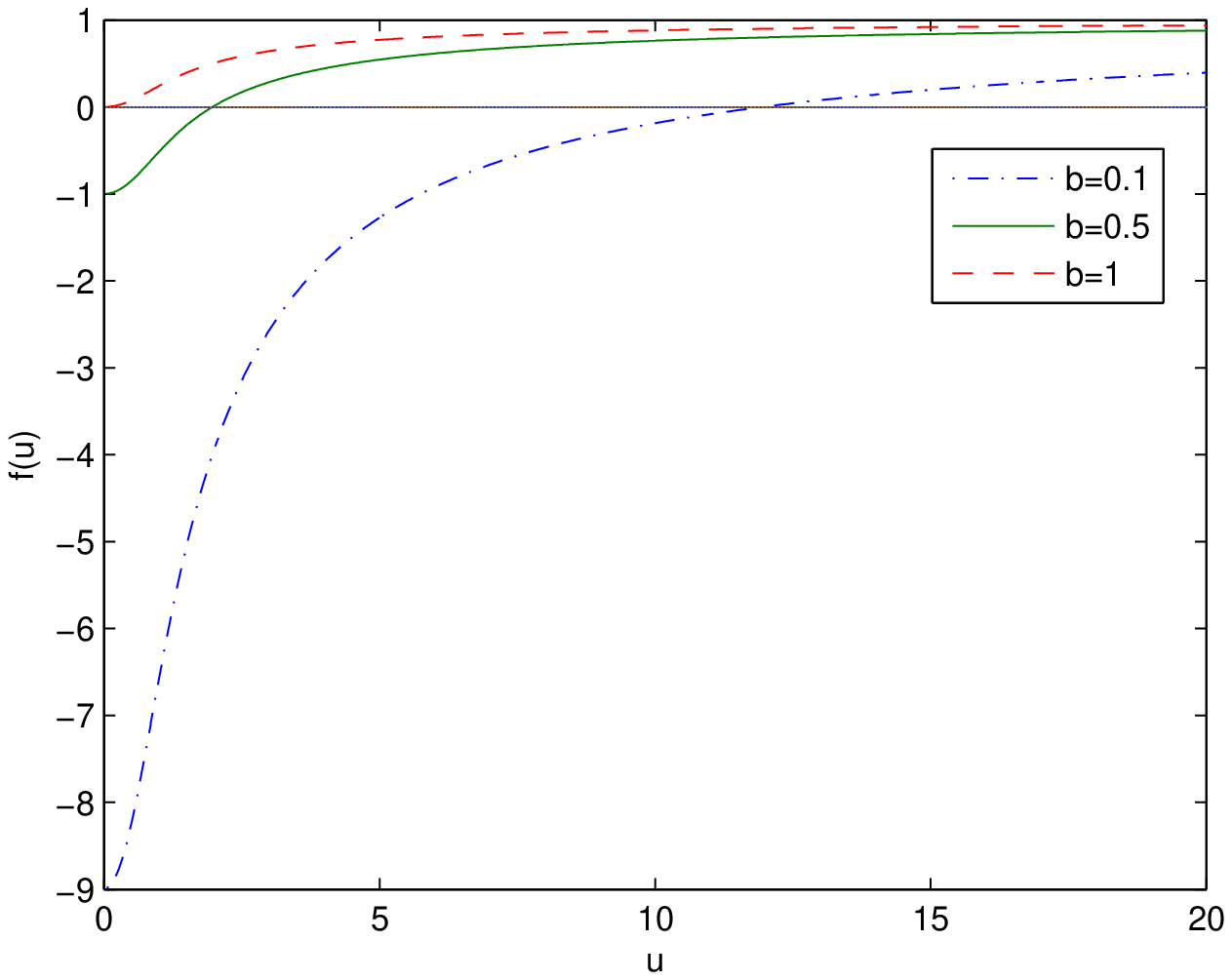}
\caption{\label{fig.2}The plot of the function $f(u)$ for $\sigma=0.5$. Dashed-dotted line corresponds to $b=0.1$, solid line corresponds to $b=0.5$ and dashed line corresponds to $b=1$.}
\end{figure}
\begin{figure}[h]
\includegraphics[height=3.0in,width=3.0in]{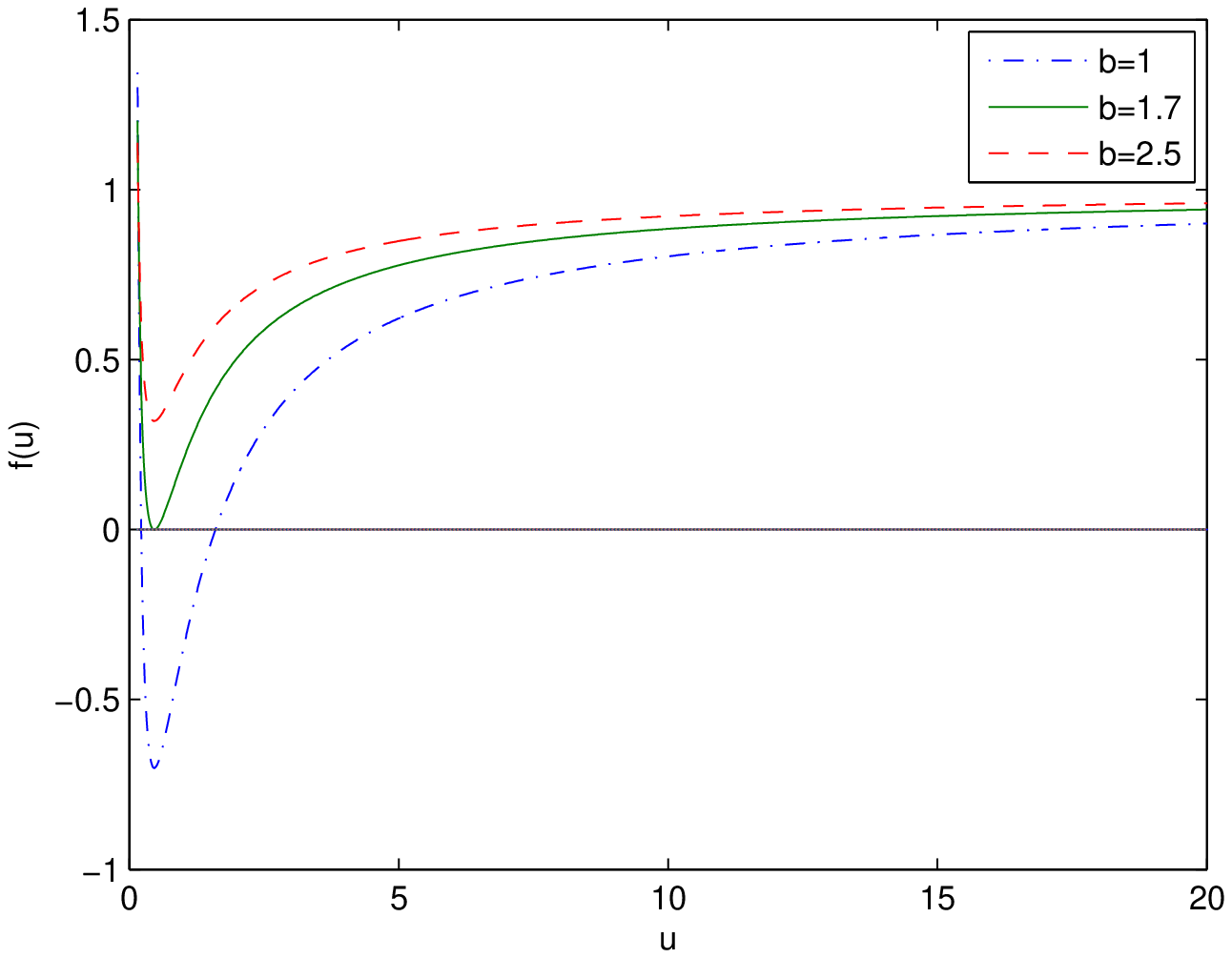}
\caption{\label{fig.3}The plot of the function $f(u)$ for $\sigma=0.6$. Dashed-dotted line corresponds to $b=1$, solid line corresponds to $b=1.7$ and dashed line corresponds to $b=2.5$.}
\end{figure}
In accordance with Fig. 1 for $\sigma=0.1$, $b=0.05$ we have two horizons, for $b=0.064$ there is one extreme horizon, and for $b=0.1$ there are not horizons (naked singularity). In addition, $f(0)=f(\infty)=1$ and, as a result, the BHs are regular. Figure 2 shows that for $\sigma=0.5$ only one event horizon exists. According to Fig. 3 there can be one, two or no horizons. The asymptotic of the hypergeometric function as $u\rightarrow 0$ is given by
\begin{equation}\label{25}
  _2F_1\left(-\frac{3}{4},-\sigma;\frac{1}{4};-\frac{1}{u^4}\right)\rightarrow
\frac{3}{(3-4\sigma)u^{4\sigma}}+\frac{4\sigma\Gamma(1/4)\Gamma(7/4-\sigma)}{(4\sigma-3)
\Gamma(1-\sigma)u^3}.
\end{equation}
Making use of Eqs. (21), (25) we obtain
\[
\lim_{u\rightarrow 0} f(u)=1-\frac{(2\sigma)^{3/4}\bar{m}_M}{bu}
\]
\begin{equation}\label{26}
-\frac{u^2}{3b}\left[\frac{3}{(3-4\sigma)u^{4\sigma}}+\frac{4\sigma\Gamma(1/4)\Gamma(7/4-\sigma)}{(4\sigma-3)
\Gamma(1-\sigma)u^3}-1\right].
\end{equation}
The asymptotic of the metric function found from  Eq. (26) and Table 1 are $f(0)=1$ for $\sigma=0.1$. For $\sigma=0.5$ and $b=0.1$ we have $f(0)=-9$, for $b=0.5$ one has $f(0)=-1$, and for $b=1$ the asymptotic is $f(0)=0$. In the case of $\sigma=0.6$ we have $f(0)=\infty$. These results are in accordance with figures 1, 2 and 3.

\section{Thermodynamics}

To study the thermal stability of charged BHs we will calculate the Hawking temperature and heat capacity. If the heat capacity is negative the BHs are locally non stable. The phase transitions take place when the Hawking temperature and heat capacity become zero or the heat capacity is singular \cite{Davies}. The Hawking temperature is given by
\begin{equation}
T_H=\frac{f'(r)|_{r_+}}{4\pi},
\label{27}
\end{equation}
where $r_+$ is the event horizon radius ($f(r_+)=0$). From Eqs. (21) and (27) we obtain the Hawking temperature
\begin{equation}
T_H=\frac{1}{4\pi}\left(\frac{2\sigma e^\gamma}{\beta Q^2}\right)^{1/4}\left(\frac{1}{u_+}-
\frac{3u_+^2\left[\left(1+1/u_+^4\right)^\sigma-1\right]}{3(2\sigma)^{3/4}\bar{m}_M +u_+^3\left[
_2F_1\left(-\frac{3}{4},-\sigma;\frac{1}{4};-\frac{1}{u_+^4}\right)-1\right]}\right).
\label{28}
\end{equation}
The plot of the Hawking temperature vs. $u_+$ is presented in Fig. 4.
\begin{figure}[h]
\includegraphics[height=3.0in,width=3.0in]{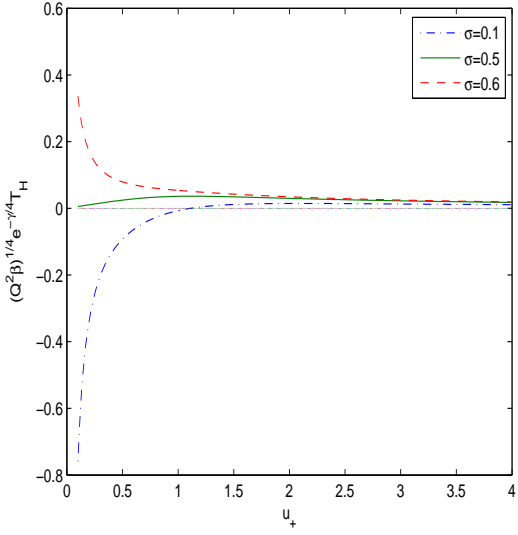}
\caption{\label{fig.4}The plot of the function $T_H\sqrt{Q}\beta^{1/4}e^{-\gamma/4}$ vs. $u_+$. Dashed-dotted line corresponds to $\sigma=0.1$, solid line corresponds to $\sigma=0.5$ and dashed line corresponds to $\sigma=0.6$.}
\end{figure}
According to Fig. 4 the Hawking temperature has a maximum at $\sigma=0.1$, $\sigma=0.5$  and the BHs are not locally stable. For $\sigma=0.6$ there is not an extremum of the Hawking temperature and a phase transition is absent. The heat capacity is given by
\begin{equation}
C_q=T_H\left(\frac{\partial S}{\partial T_H}\right)_q=\frac{\partial M}{\partial T_H}=\frac{\partial M/\partial u_+}{\partial T_H/\partial u_+}.
\label{29}
\end{equation}
Equation (29) shows that the heat capacity possesses a singularity if the Hawking temperature has an extremum.
In accordance with Fig. 4 there is a maximum of the Hawking temperature for some parameters $\sigma$ and, therefore, phase transitions take place. From Eqs. (16) and (28) we obtain
\[
\frac{\partial M}{\partial u_+}=\frac{Q^{3/2}u_+^2}{(2\sigma)^{3/4}e^{3\gamma/4}\beta^{1/4}}\left[\left(1+\frac{1}{u_+^4}\right)^\sigma-1\right],
\]
\[
\frac{\partial T_H}{\partial u_+}=\frac{1}{4\pi}\left(\frac{2\sigma e^{\gamma}}{\beta Q^2}\right)^{1/4}
\biggl\{ -\frac{1}{u_+^2}-\frac{6u_+\left[\left(1+\frac{1}{u_+^4}\right)^\sigma-1\right]-\frac{12\sigma}{u_+^3}\left(1+\frac{1}{u_+^4}
\right)^{\sigma-1}}{3\bar{m}(2\sigma)^{3/4}+u_+^3\left[_2F_1\left(-\frac{3}{4},-\sigma;\frac{1}{4};-\frac{1}{u_+^4}\right)-1\right]}
\]
\begin{equation}
+\frac{9u_+^4\left[\left(1+\frac{1}{u_+^4}\right)^\sigma-1\right]^2}{\left(3\bar{m}(2\sigma)^{3/4}
+u_+^3\left[_2F_1\left(-\frac{3}{4},-\sigma;\frac{1}{4};-\frac{1}{u_+^4}\right)-1\right]\right)^2}\biggr\}.
\label{30}
\end{equation}
Making use of Eqs. (29) and (30) we plot the heat capacity vs. $u_+$.
\begin{figure}[h]
\includegraphics[height=3.0in,width=3.0in]{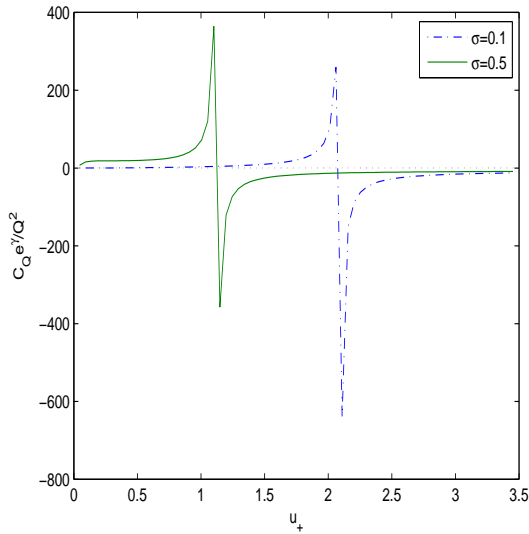}
\caption{\label{fig.5}The plot of the function $C_Qe^{\gamma}/Q^2$ vs. $u_+$. Dashed-dotted line corresponds to $\sigma=0.1$ and solid line corresponds to $\sigma=0.5$.}
\end{figure}
Figure 5 shows that indeed the heat capacity diverges in the points where the Hawking temperature has a maximum and phase transitions occur.
There are areas with positive and negative heat capacities which are separated by discontinuity points. The negative capacity corresponds to the BH unstable state and the early stage of the thermodynamics process. The positive heat capacity belongs to the BH stable state and the late stage of the thermodynamics process. If the parameter $\sigma$ smaller, the second-order BH phase transition takes place at the larger value of the horizon radius $r_+$.
In accordance with Eq. (29) at a point $u_+=u_1$ the Hawking temperature and heat capacity are zero and a first-order phase transition happens and the BH remnant is formed. In this point $u_1$ the BH mass is not zero but the Hawking temperature and the heat capacity vanish.  When the maximum of the Hawking temperature occurs (at $u_+=u_2$) the heat capacity diverges and the second-order phase transition takes place.
The values of $u_1$ and $u_2$ for some parameters $\sigma$ are given in Table 2.
\begin{table}[ht]
\caption{The values of $u_1$ and $u_2$}
\centering
\begin{tabular}{c c c c c c c c c c }\\[1ex]
% centered columns
\hline
$\sigma$ & 0.05 & 0.1 & 0.15 & 0.2 & 0.25 & 0.3 & 0.35 &0.4 & 0.45   \\[0.5ex]
\hline
$u_1$ & 1.157 & 1.106 & 1.058 & 1.002 & 0.942 & 0.866 & 0.776 & 0.661 & 0.494 \\[0.5ex]
\hline
$u_2$ & 2.164 & 2.089 & 2.018 & 1.937 & 1.851 & 1.751 & 1.637 & 1.505 & 1.345  \\[0.5ex]
\hline
\end{tabular}
\end{table}
At the range $u_2>u_+>u_1$ BHs are locally stable and at $u_+>u_2$ the BHs become unstable. At $0<u_+<u_1$ one has the BH remnant.

 \section{Conclusion}

We have studied GenModMax model of nonlinear electrodynamics with four independent parameters $\beta$, $\lambda$, $\gamma$ and $\sigma$. At some parameters the model becomes ModMax electrodynamics, BI electrodynamics or BI-type electrodynamics. The correspondence principle holds and at weak fields the model is converted into Maxwell's electrodynamics. The singularity of the electric field at the origin of point-like charged particles is absent at $\sigma<1$. We investigate GenModMax model coupled to general relativity.
The mass and metric functions of the magnetized BH are obtained. It was shown that the BH can have one, two or no horizons.
It is worth noting that for $\sigma=0.1$ one has $f(0)=f(\infty)=1$ and, therefore, the regular BH solution holds.
We obtained the asymptotic of the metric and mass functions as $r\rightarrow\infty$ and $r\rightarrow 0$, and
corrections to the Reissner--Nordstr\"{o}m solution. The Hawking temperature and heat capacity of BHs were calculated and it is shown that BHs are not locally stable for $0<\sigma<0.5$ and phase transitions take place where the Hawking temperature has a maximum and the heat capacity diverges.


\begin{thebibliography}{99}

\bibitem{Born} M. Born and L. Infeld, Proc. Royal Soc. (London) A \textbf{144}, 425 (1934).
%Foundations of the New Field Theory,

\bibitem{Fradkin} E. S. Fradkin and A. Tseytlin, Phys. Lett B \textbf{163}, 123 (1985).

\bibitem{Tseytlin} A. Tseytlin, Nucl. Phys. B \textbf{276}, 391 (1985).

\bibitem{Shabad} D. M. Gitman and A. E. Shabad, Eur. Phys. J. C \textbf{74}, 3186 (2014), arXiv:1410.2097 [hep-th].

\bibitem{Shabad1} C. V. Costa, D. M. Gitman and A. E. Shabad, Phys. Scripta \textbf{90}, 074012 (2015), arXiv:1312.0447  [hep-th].
	
\bibitem{Kr1} S. I. Kruglov, Ann. Phys. (Berlin) \textbf{527}, 397 (2015), arXiv:1410.7633 [physics.gen-ph].

\bibitem{Kr2} S. I. Kruglov, Commun. Theor. Phys. \textbf{66}, 59 (2016), arXiv:1511.03303 [hep-ph].

\bibitem{Kr3} S. I. Kruglov,  Ann. Phys. \textbf{353}, 299 (2015), arXiv:1410.0351 [physics.gen-ph].

% Phys. Lett. A \textbf{\textbf{379}}, 623 (2015) (arXiv:1504.03535).

\bibitem{Heisenberg} W. Heisenberg and H. Euler, Z. Physik, \textbf{98}, 714 (1936), arXiv:physics/0605038.

\bibitem{Schwinger} J. Schwinger, Phys. Rev. \textbf{82}, 664 (1951).

\bibitem{Adler} S. L. Adler, Ann. Phys. (N.Y.) \textbf{67}, 599 (1971).

\bibitem{Bandos} I. Bandos, K. Lechner, D. Sorokin, and P. Townsend,  Phys. Rev. D \textbf{102}, 121703 (2020), arXiv:2007.09092 [hep-th].
%A non-linear duality-invariant conformal extension of Maxwell’s equations,.

\bibitem{Kosyakov} B. P. Kosyakov, Phys. Lett. B \textbf{810}, 135840 (2020), arXiv:2007.13878 [hep-th].
\bibitem{Habib}Z. Amirabi, S. Habib Mazharimousavi, Eur. Phys. J. C \textbf{81}, 207 (2021), arXiv:2012.07443 [gr-qc].
%Black-hole solution in nonlinear electrodynamics with the maximum allowable symmetries e-Print:
\bibitem{Bandos1}I. Bandos, , K. Lechner, D. Sorokin, and P. K. Townsend, JHEP \textbf{03}, 022 (2021), arXiv:2012.09286 [hep-th].
%On p-form gauge theories and their conformal limits• e-Print:
\bibitem{Townsend} I. Bandos, K. Lechner, D. Sorokin, and P. Townsend, ModMax meets Susy, arXiv:2106.07547 [hep-th].
\bibitem{Flores}D. Flores-Alfonso, B. A. Gonzalez-Morales, R. Linares, and M. Maceda, Phys. Lett. B \textbf{812}, 136011 (2021), arXiv:2011.10836 [gr-qc].
%Black holes and gravitational waves sourced by non-linear duality rotation-invariant conformal electromagnetic matter,.
\bibitem{Kubiznak}A. Ballon Bordo, D. Kubiznak, and T. R. Perche, Phys. Lett. B \textbf{817}, 136312 (2021), arXiv:2011.13398 [hep-th].
\bibitem{Kruglov} S. I. Kruglov, Phys. Lett. B \textbf{822}, 136633 (2021), arXiv:2108.08250 [physics.gen-ph].
\bibitem{Kruglov0}S. I. Kruglov, Mod. Phys. Lett. A\textbf{ 32}, 1750201 (2017), arXiv:1612.04195 [physics.gen-ph].
%Notes on Born–Infeld-type electrodynamics• e-Print:
\bibitem{Kruglov1} S. I. Kruglov, Ann. Phys. \textbf{383}, 550 (2017),  Ann. Phys. \textbf{434}, 168625 (2021) (Corrigendum), arXiv:1707.04495 [gr-qc].
%Born–Infeld-type electrodynamics and magnetic black holes• e-Print:
\bibitem{Kruglov2} S. I. Kruglov, J. Phys. A \textbf{43}, 375402 (2010), arXiv:0909.1032 [hep-th].
% On generalized Born-Infeld electrodynamics• e-Print:
\bibitem{Bardeen} J. M. Bardeen, in Proc. Int. Conf. GR5, Tbilisi, p. 174, 1968.

%\bibitem{Oliveira} H. P. de Oliveira, Class. Quant. Grav. \textbf{11}, 1469 (1994).
%\bibitem{Soleng} H. H. Soleng, Phys. Rev. D \textbf{52}, 6178 (1995) (arXiv:hep-th/9509033).

\bibitem{Ayon1} E. Ay\'{o}n-Beato, A. Gar\'{c}ia, Phys. Rev. Lett.  \textbf{80}, 5056 (1998),
arXiv:gr-qc/9911046 [gr-qc].

\bibitem{Bronnikov} K. A. Bronnikov, Phys. Rev. D \textbf{63}, 044005 (2001).

\bibitem{Breton} N. Breton, Phys. Rev. D \textbf{67}, 124004 (2003), arXiv:hep-th/0301254.

\bibitem{Hayward} S. A. Hayward, Phys. Rev. Lett. \textbf{96}, 031103 (2006), arXiv:gr-qc/0506126.

\bibitem{Breton1} N. Breton and R. Garcia-Salcedo, Nonlinear Electrodynamics and black holes, arXiv:hep-th/0702008.

\bibitem{Lemos} J. P. S. Lemos and V. T. Zanchin, Phys. Rev. D \textbf{83}, 124005 (2011),
arXiv:1104.4790 [gr-qc].

\bibitem{Lemos1} A. Flachi and J. P. S. Lemos, Phys. Rev. D \textbf{87}, 024034 (2013),
arXiv:1211.6212 [gr-qc].

%\bibitem{Hendi} S. H. Hendi, Ann. Phys. \textbf{333}, 282 (2013) (arXiv:1405.5359 [gr-qc]).

\bibitem{Balart} L. Balart and E. C. Vagenas, Phys. Rev. D \textbf{90}, 124045 (2014), arXiv:1408.0306 [gr-qc].

\bibitem{Kruglov3} S. I. Kruglov, Phys. Rev. D \textbf{94}, 044026 (2016), arXiv:1608.04275 [gr-qc].

\bibitem{Kruglov4} S. I. Kruglov,  Europhys. Lett. \textbf{115}, 60006 (2016), arXiv:1611.02963 [physics.gen-ph].

\bibitem{Kruglov5} S. I. Kruglov, Ann. Phys. (Berlin) \textbf{528}, 588 (2016), arXiv:1607.07726 [gr-qc].

\bibitem{Frolov} V. P. Frolov, Phys. Rev. D \textbf{94}, 104056 (2016), arXiv:1609.01758 [gr-qc].

\bibitem{Pellicer} R. Pellicer and R. J. Torrence, J. Math. Phys. \textbf{10}, 1718 (1969).

\bibitem{Fan} Zhong-Ying Fan and Xiaobao Wang, Phys. Rev. D \textbf{94},124027 (2016),	arXiv:1610.02636 [gr-qc].

\bibitem{Bronnikov1} K. A. Bronnikov, Phys. Rev. D \textbf{96}, 128501 (2017),  arXiv:1712.04342 [gr-qc].

\bibitem{Kruglov6}S. I. Kruglov, Mod. Phys. Lett. A \textbf{32},  1750092 (2017), arXiv:1705.08745 [physics.gen-ph].

\bibitem{Bateman} H. Bateman and A. Erdelyi, \textit{Higher Transcendental Functions}, Vol. 1, Mc. Graw-Hill Book Company, Inc., 1953.

\bibitem{Davies} P. C. W. Davies, Rep. Prog. Phys. \textbf{41}, 1313 (1978).

%\bibitem{Kruglov6} S. I. Kruglov, Ann. Phys. (Berlin) \textbf{528}, 588 (2016).
%(arXiv:1607.07726 [gr-qc]).

%\bibitem{Kruglov8} S. I. Kruglov, Phys. Rev. D \textbf{94}, 044026 (2016) (arXiv:1608.04275 [gr-qc]).


\end{thebibliography}
\end{document}